\documentstyle[epsfig]{aipproc}

\begin{document}
\title{Phase Resolved Spectroscopy of Burst Oscillations: Searching for 
Rotational Doppler Shifts}

\author{Tod E. Strohmayer}
\address{Laboratory for High Energy Astrophysics \\ NASA's Goddard Space Flight
Center \\ Mail Code 662, Greenbelt, MD 20771}

\maketitle

\begin{abstract}
X-ray brightness oscillations with frequencies from 300 - 600 Hz have
been observed in six low mass X-ray binary (LMXB) bursters with the 
Rossi X-ray Timing Explorer (RXTE). These oscillations likely result from 
spin modulation of a hot region on the stellar surface produced by 
thermonuclear burning. If this hypothesis is correct the rotational velocity 
of the stellar surface, $\approx 0.1 \; c$, will introduce a pulse phase
dependent Doppler shift such that the rising edge of a pulse should be
harder (blue shifted) than the trailing edge (red shifted). Detection of this
effect would both provide further compelling evidence for the spin modulation
hypothesis as well as providing new observational techniques with which to
constrain the masses and radii of neutron stars. In this work I present 
results of an attempt to search for such Doppler shifts.
\end{abstract}

\section*{Introduction}

A rotating hot spot (or spots) seems to be the most simple, consistent scenario
proposed to date to explain the presence of large amplitude modulations of
the X-ray brightness during thermonuclear bursts. The presence of large
amplitudes at burst onset combined with spectral evidence for localized 
X-ray emission supports this hypothesis \cite{szs}. Additional evidence
for spin modulation is provided by the fact that the oscillation frequencies
are stable over year timescales and within bursts the oscillations are
highly coherent \cite{s98b,sm}

Detailed studies of the burst oscillations hold great promise for providing new
insights into a variety of physics issues related to the structure and 
evolution of neutron stars. For example, within the context of the rotating 
hot spot model it is possible to determine constraints on the mass and radius 
of the neutron star from measurements of the maximum observed modulation 
amplitudes during X-ray bursts as well as the harmonic content of the pulses
\cite{ml,s98a}. Phase resolved X-ray spectroscopy of the burst oscillations
also holds great promise, and could yield methods to constrain the radii of
neutron stars, a quantity which is extremely difficult to infer on its own.
For example, a 10 km radius neutron star spinning at 400 Hz has a surface 
velocity of $v_{spin}/c \le 2\pi \nu_{spin} R/c  \approx 0.084$ at the 
rotational equator. This motion of the hot spot produces a Doppler shift of 
magnitude $\Delta E / E \approx v_{spin}/c$, thus the observed spectrum is a 
function of pulse phase \cite{cs}. Measurement of this phase dependent Doppler 
shift would provide further compelling evidence supporting the spin modulation 
model and also a means of constraining the neutron star radius, since for a 
known spin frequency the velocity, and thus magnitude of the Doppler shift, is 
proportional to the stellar radius. In addition, both the magnitude of the 
Doppler shift and the amplitude of spin modulation decrease as the lattitude
of the hot spot increases (Both approaching zero when the spot is located at 
the rotational pole). Detection of this correlation in a large sample of bursts
would provide definitive proof of the spin modulation hypothesis.

\section*{Phase Resolved Spectroscopy}

Detailed searches for a Doppler shift signature are just beginning to be 
carried out. Studies using the oscillations in single bursts have shown that 
4-5 \% modulations of the fitted black body temperature are easily detected 
in the tails of bursts \cite{ssz}, and are consistent with the idea that a 
temperature gradient is present on the stellar surface, which when rotated 
produces the flux modulations. Phase lag studies in a burst from Aql X-1 
indicates that softer photons lag higher energy photons in a manner which is 
qualitatively similar to that expected from a rotating hot spot \cite{ford}.

\begin{figure}[b!] 
\centerline{\epsfig{file=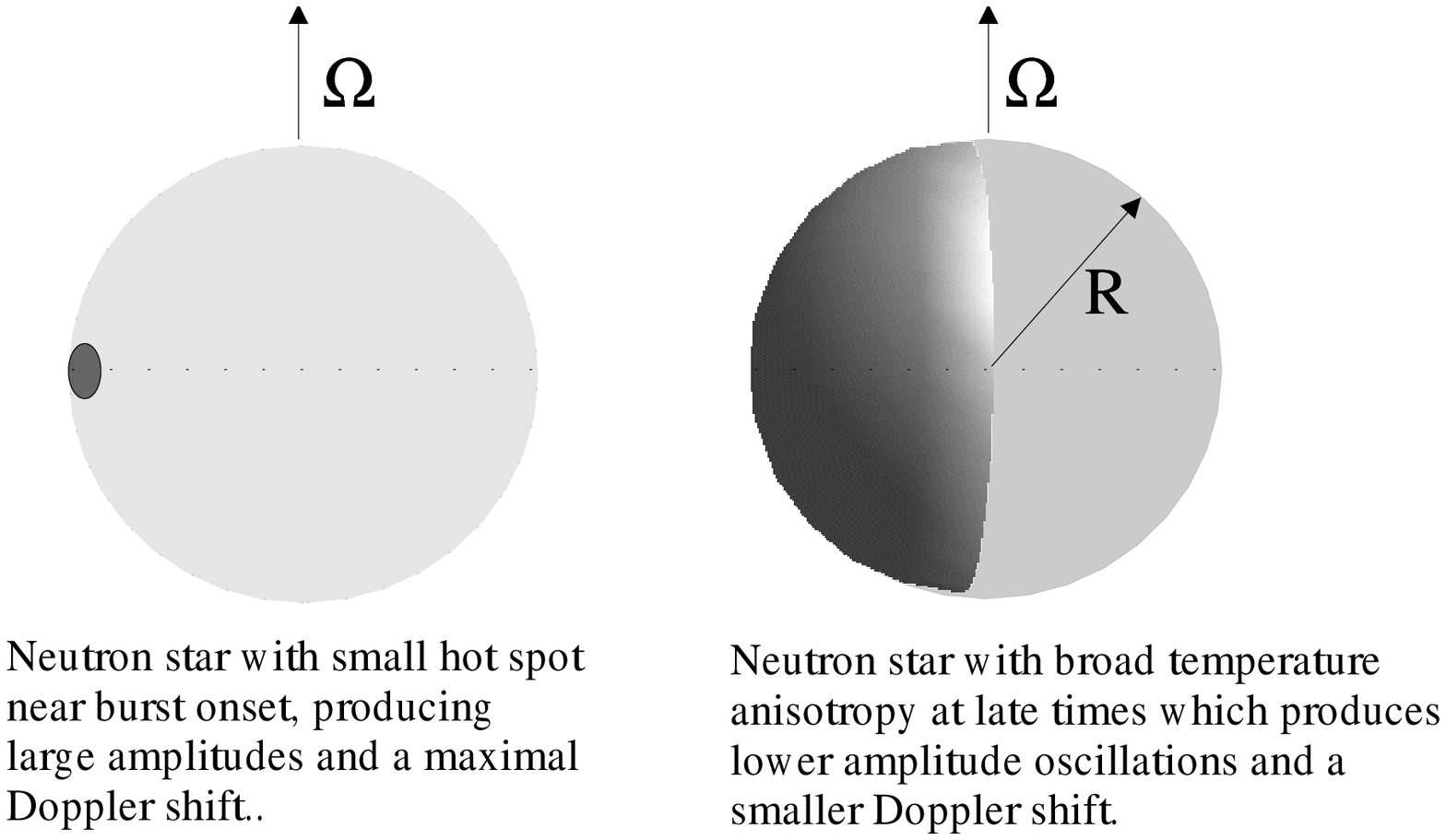,height=3.0in,width=5.5in}}
\vspace{5pt}
\caption{}
\label{fig1}
\end{figure}

There are several ways to address the issue of how best to
search for Doppler shifts. Near burst onset the oscillation amplitude is 
high and the hot spot is well localized. If it were possible to examine the
oscillations with exquisite detail nearer and nearer to the burst onset then
the situation should approach that of a point-like hot spot on the star. 
A point spot located on the rotational equator produces the largest modulation
amplitude as well as a maximal Doppler shift. Thus, in principle this would be
an ideal interval to examine, however, the difficulty is that the burst flux is
still weak near onset and the interval during which the spot is small (to 
accumulate a spectrum) is short. This means that the desired signal to examine 
is very weak. An alternative is to examine pulse trains in the decaying tails
of bursts \cite{sm}. In this case the oscillation interval is much longer, the 
burst flux is still substantial and there are many more photons available. 
However, the oscillation amplitude is lower and at late times in bursts we 
expect that a well localized spot is probably not present. More likely the 
modulations are caused by a broad temperature anisotropy over the stellar 
surface (see Figure \ref{fig1}. This means that the observed Doppler shift 
represents an integral over the rotating surface of the star, much of which is 
moving with a lower line of sight velocity than the maximum at the rotational 
equator. Thus the Doppler shift will be weaker than what we might expect near 
burst onset. To some extent the lack of signal in a single burst can be offset
by summing data from several bursts coherently \cite{sm,m}. 

\section*{Results}

To search for a Doppler shift I have examined the phase resolved spectrum of 
4 bursts with 330 Hz oscillations from 4U 1702-429 \cite{mss}. Intervals in
the tails of these bursts were analysed first since these have much more
signal in the pulsations than intervals near onset. For each burst I fit an 
exponential model of the frequency drift, as in \cite{sm}, and then used the 
phase information for each X-ray event, based on the best model, to accumulate 
spectra in 12 phase bins. The data were recorded in an event mode which 
provides the arrival time and energy channel of each X-ray. The data mode has
64 energy channels in the $\approx$ 2 - 100 keV PCA band. 
As a measure of spectral hardness versus pulse phase I used the mean PCA energy
bin for each phase interval. In principle one could also fit the accumulated 
spectra with a black body function and use the color temperature as a hardness
measure, however, the spectra are accumulated over an interval long compared to
the burst cooling time so that the spectra are not well fit by black body 
functions with a constant temperature. In each burst there is a strong
modulation of the mean PCA energy channel with pulse phase, with the peak of 
the phase folded lightcurve having the highest mean channel. This result is
similar to that found for oscillations at 580 Hz in a burst from 4U 1636-53
\cite{ssz}, that is, the peak of the modulations are harder than the minima.

\begin{figure}[b!] 
\centerline{\epsfig{file=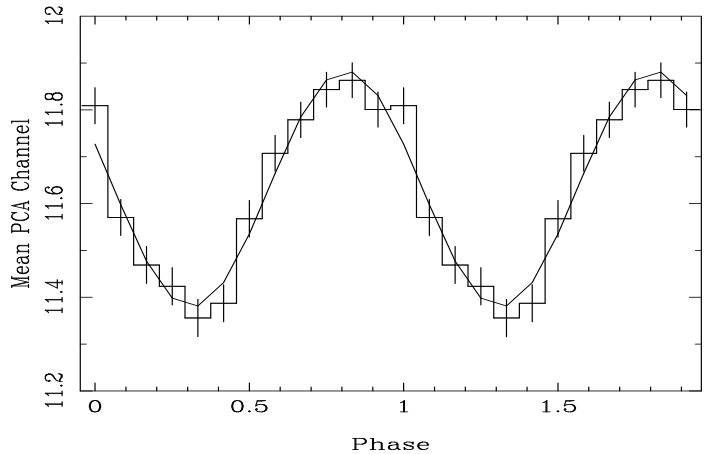,height=2.5in,width=5.5in}}
\vspace{10pt}
\caption{Variation of the mean PCA channel from spectra accumulated as a 
function of pulse phase in 4U 1702-429. Data from 4 bursts were added 
coherently in order to increase the signal to noise ratio. There is no strong
evidence for a Doppler shift induced asymmetry.}
\label{fig2}
\end{figure}

The single burst data show obvious modulations of the mean PCA channel with 
pulse phase, however, they do not show evidence for any asymmetry such as 
could be introduced by the rotational Doppler shift. This is mainly due to the
lack of sufficient signal in a single burst. In an attempt to overcome these
limitations I co-added the data for all 4 bursts in phase \cite{sm}. This 
amounts to finding a phase offset, relative to one burst, which maximizes the
pulsation signal of the sum. I then computed the mean PCA channel for the 
summed spectra in the same way as for the individual bursts. The results are
given in figure \ref{fig2}. There is a strong 4 \% modulation of the mean
PCA channel in the combined data. As a simple test for asymmetry I fit a sine
function to the mean PCA channel data and find that the data are adequately 
described by a simple sine wave, so we do not find any strong evidence for a
Doppler shift in the combined data. It still remains to conduct more 
sophisticated tests for asymmetry on these data, and more bursts will be added 
as new data becomes available.  

\end{document}